\documentstyle[twoside,fleqn,espcrc2]{article}

\newcommand{\AmS}{{\protect\the\textfont2
  A\kern-.1667em\lower.5ex\hbox{M}\kern-.125emS}}

\title{A High Statistics Lattice Calculation of Quark Masses 
       with a Non-Perturbative Renormalization Procedure.}

\author{L.~Giusti\address{Scuola Normale Superiore and INFN Sezione di Pisa, 
        P.~zza dei Cavalieri 7, I-56100 Pisa Italy}%
        \thanks{Work done in collaboration with V.~Gim\'enez, F.~Rapuano and
M.~Talevi} }

\begin{document}

\begin{abstract}
\vspace{-.3cm}
We present results of a high statistics study  ($O(2000)$ configurations) of the
quark masses in the  $\overline{MS}$ scheme from Lattice QCD in the quenched
approximation at $\beta=6.0$, $\beta=6.2$ and $\beta=6.4$ using both the 
Wilson and the tree-level improved SW-Clover fermion  action. We extract 
quark masses from the meson spectroscopy  and from the axial Ward Identity 
using non-perturbative values of the renormalization constants. We compare 
the results obtained with the two methods and we study the $O(a)$ 
dependence of the quark masses for both actions. Our best results are 
$m_s^{\overline{MS}}(2\; GeV)=(123\pm 4\pm 15)\; MeV$ and
$m_c^{\overline{MS}}(2\; GeV)=(1525\pm 40\pm 100)\; MeV$.              
\end{abstract}

\maketitle
\vspace{-.5cm}
\section{INTRODUCTION}
Quark masses are fundamental parameters of the QCD Lagrangian that cannot be
measured directly in the experiments, since free quarks are not physical
states, and that cannot be predicted by QCD using only theoretical 
considerations. Up to now QCD sum rules and lattice simulations are 
the only non-perturbative
techniques able to determine the absolute values of the light quark masses.
Lattice technique does not require any additional model parameter and 
is the only procedure that can be systematically improved.       
\vspace{-.3 cm}
\section{QUARK MASSES}
The usual on-shell mass definition cannot be used for quark masses and 
their values depend on the theoretical definition adopted. In the following we
will give our final results for the quark masses defined in the 
$\overline{MS}$ scheme.
$m^{\overline{MS}}(\mu)$ is defined by the perturbative expansion of the 
quark propagator renormalized with the $\overline{MS}$ prescription
\cite{guido} and it depends on the renormalization scale $\mu$. The symbols 
used in the following are fully defined in \cite{guido,finale}.\\ 
The ``bare'' lattice quark mass can be determined directly from lattice
simulations by fixing the mass of a hadron containing
a quark with the same flavour to its experimental value.\\ 
\begin{table*}[hbt]
\setlength{\tabcolsep}{.2pc}
\newlength{\digitwidth} \settowidth{\digitwidth}{\rm 0}
\catcode`?=\active \def?{\kern\digitwidth}
\label{tab:latparams}
\caption{Summary of the parameters of the runs analyzed in this work.} 
\begin{tabular}{lllllllllll}
\hline         
\multicolumn{11}{c}{Matrix Elements} \\
        &C60a&C60b&C60c&C60d&W60&C62a&W62a&W62b&C64&W64\\
\hline
$\beta$ &6.0 &6.0 &6.0 &6.0 &6.0&6.2 &6.2 &6.2 &6.4&6.4\\
Action  &SW  &SW  &SW  &SW  &Wil&SW  &Wil &Wil &SW &Wil\\
\# Confs&490 &600 &200 &200 &320&250 &250 &110 &400&400\\
Volume  &$18^3\times 64$&$24^3\times 40$&$18^3\times 32$&
         $16^3\times 32$&$18^3\times 64$&$24^3\times 64$&
         $24^3\times 64$&$24^3\times 64$&$24^3\times 64$&
         $24^3\times 64$\\
\hline
\multicolumn{11}{c}{   } \\
\hline
\multicolumn{11}{c}{Renormalization Constants} \\
\multicolumn{3}{c}{ } &C60Z&W60Z&C62Z&W62Z&C64Z&W64Z&\multicolumn{2}{c}{ }\\
\hline
\multicolumn{3}{c}{$\beta$} &6.0&6.0&6.2&6.2&6.4&6.4&\multicolumn{2}{c}{ }\\
\multicolumn{3}{c}{Action } &SW &Wil&SW &Wil&SW &Wil&\multicolumn{2}{c}{ }\\
\multicolumn{3}{c}{\# Confs}&100&100&180&100&60 &60 &\multicolumn{2}{c}{ }\\
\multicolumn{3}{c}{Volume } &$16^3\times 32$&$16^3\times 32$&
                             $16^3\times 32$&$16^3\times 32$&
                             $24^3\times 32$&$24^3\times 32$&
                             \multicolumn{2}{c}{ }\\
\hline
\end{tabular}
\end{table*}
Quark masses can also be extracted from the axial Ward Identity. Close to
the chiral limit and neglecting terms of $O(a)$, the Ward Identity can be
written as
\begin{equation}
Z_A \langle\alpha|\partial^\mu A_\mu^a|\beta\rangle = 2 (m - m_c) 
\frac{Z_P}{Z_S}\langle\alpha| P^a|\beta\rangle\; . 
\end{equation}
where $Z_A$, $Z_P$ and $Z_S$ are the renormalization constants of the axial,
pseudoscalar and scalar densities and $m_c$ is defined in 
\cite{guido,bochi}. 
The standard perturbative approach uses the lattice and the continuum 
perturbation theory to connect the ``bare'' lattice quark mass to 
the $m^{\overline{MS}}(\mu)$ \cite{guido,finale}.
The scale $1/a$, where $a$ is the lattice spacing, of our 
simulations is $a^{-1}\simeq 2-4$ GeV. At these scales we expect small 
non-perturbative effects. However the
``tadpole'' diagrams, which are present in the lattice perturbation
theory, can give raise to large perturbative corrections and then to large
uncertainties in the matching procedure at values of 
$\beta=6/g_L^2=6.0-6.4$.\\
The non-perturbative renormalization (NP) techniques eliminate these 
uncertainties \cite{rinnp,luescher}. $Z_A$, $Z_P$ and $Z_S$ can be 
calculated by imposing the renormalization conditions, described in the next 
section, on the 
quark states of momentum $p^2=\mu^2$ and in the Landau gauge\cite{rinnp}.
The quark mass in the $\overline{MS}$ scheme is then defined as
\begin{eqnarray}
m^{\overline{MS}}(\mu) & = & U_m^{\overline{MS}}(\mu,\mu')
\left[1+ \frac{\alpha_s(\mu')}{4\pi}C^{LAN}_m\right]\cdot \nonumber\\
& & \hspace{-.7 cm}\frac{Z_A}{Z_P^{RI}(\mu' a)}\rho(a)
\end{eqnarray}
where
\begin{eqnarray}
& &\hspace{-.7 cm}U_m^{\overline{MS}}(\mu,\mu') =  (\frac{\alpha_s(\mu)}
                       {\alpha_s(\mu')})^{\gamma^{(0)}/2\beta_0}\left[
      1+ \right.\nonumber\\
& &  + \left.\frac{\alpha_s(\mu)-\alpha_s(\mu')}{4\pi}
(\frac{\gamma^{(1)}}{2\beta_0}-\frac{\gamma^{(0)}\beta_1}{2\beta_0^2})\right]
\end{eqnarray}                                     
and, for large time separations,
\begin{equation}
\rho(a) = \frac{1}{2}M_{PS}\frac{\langle A_0(\tau) P(0) \rangle}
     {\langle P(\tau) P(0)\rangle}\; .
\end{equation}
\setlength{\tabcolsep}{.35pc}
\begin{table}
\caption{Quark Masses from the spectroscopy in MeV. $\overline{MS}$ masses 
are at $\mu=2$ GeV.}
\label{tab:qmas_spect}

\begin{tabular}{lllll}
\hline
 Run     & 
         $m_s(a)$ & $m_s^{\overline{MS}}$ & 
         $m_c(a)$ & $m_c^{\overline{MS}}$ \\
\hline
C60a   & 83(2) & 118(7)  &    -     &  -  \\
C60b   & 81(2) & 117(7)  &    -     &  -  \\
C60c   & 83(3) & 119(7)  &    -     &  -  \\
C60d   & 79(3) & 113(7)  &    -     &  -  \\
W60    & 98(2) & 133(16) & 1335(31) & 1612(316)  \\
C62a   & 83(4) & 120(9)  & 1106(49) & 1470(170)  \\
W62a   & 93(3) & 131(15) & 1205(24) & 1553(249)  \\
W62b   & 92(4) & 129(16) & 1206(32) & 1560(257)  \\
W64    & 82(3) & 120(16) &    -     &  -  \\
C64    & 69(3) & 103(9)  &    -     &  -  \\
\hline
\end{tabular}
\end{table}
This procedure uses only the continuum perturbation theory to connect the 
quark mass from the $RI$-scheme to $\overline{MS}$. The continuum
perturbation theory is used at scales $\mu\simeq 2-4$ GeV large enough to 
avoid non-perturbative effects or higher order corrections.\\
The non-perturbative approach can be applied directly to the unquenched 
case \cite{tchi}.  
\vspace{-.3cm}      
\section{NP RENORMALIZATION}\label{non-pert}
\begin{figure}[t]   
       \setlength{\unitlength}{1truecm}
       \begin{picture}(4.0,4.0)
         {\includegraphics{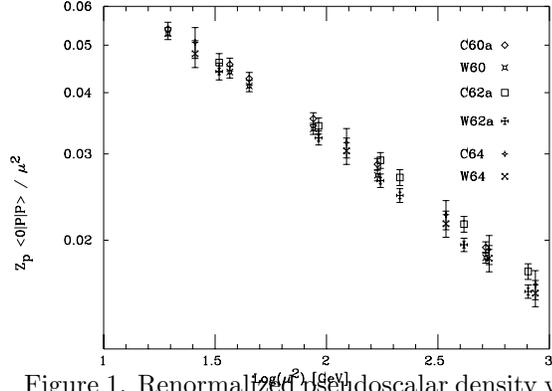}}
       \end{picture}
\caption{Renormalized pseudoscalar density versus $\log(\mu^2)$ for the
runs analysed.}
\protect\label{fig.1}
\end{figure}

For a generic two-quark operator $O_{\Gamma}(a) = \bar{\psi}\Gamma\psi$, we
define the renormalized operator 
$O_{\Gamma}(\mu) = Z_{O_\Gamma}^{RI}O_\Gamma(a)$ by introducing the
renormalized constant $Z_{O_\Gamma}^{RI}$ calculated imposing the
renormalization condition \cite{rinnp} 
\[
Z_{O_\Gamma}^{RI}(\mu a)Z^{-1}_{\psi}(\mu
a)\Gamma_{O_\Gamma}(pa)|_{p^2=\mu^2} =1 \; .
\]
$\Gamma_{O_\Gamma}(pa)$ is the forward amputated green function of the bare
operator $O_{\Gamma}(a)$ on off-shell quark states with $p^2=\mu^2$ in the
Landau gauge. For $Z_{\psi}$ we have used a definition inspired by the Ward
Identities   
\[
Z_\psi= \frac{Tr\left(\sum_{\lambda=1,4}\gamma_\lambda \sin(p_\lambda a
)S^{-1}(pa)\right)}{48 i \sum_{\lambda=1,4}\sin^2(p_\lambda a)}|_
{p^2=\mu^2}\; , 
\]
where $S(pa)$ is the quark propagator. These renormalization conditions 
satisfy the continuum Ward Identity, i.e. $Z_P/Z_S =
Z_P^{RI}/Z_S^{RI}$.\\
\setlength{\tabcolsep}{.35pc}
\begin{table}
\caption{Quark Masses from the Ward Identity in MeV. $\overline{MS}$ masses 
are at $\mu=2$ GeV.}                                  
\label{tab:qmas_wi}
\begin{tabular}{lllll}
\hline
 Run    & 
         $\rho_s(a)$ & $m_s^{\overline{MS}}$ & 
         $\rho_c(a)$ & $m_c^{\overline{MS}}$ \\
\hline
C60a   & 57(2) & 129(6) & -  &  -  \\
C60b   & 55(1) & 125(5) & -  &  -  \\
C60c   & 59(2) & 132(6) & -  &  -  \\
C60d   & 55(2) & 127(6) & -  &  -  \\
W60    & 77(2) & 126(4) & 956(20)  & 1557(52)  \\
C62a   & 60(4) & 127(9) & 747(25)  & 1572(71)  \\
W62a   & 75(3) & 120(6) & 936(16)  & 1502(47)  \\
W62b   & 71(3) & 115(6) & 927(22)  & 1491(55)  \\
W64    & 69(4) & 107(7) &  - &  -  \\
C64    & 55(4) & 106(8) &  - &  -  \\
\hline
\end{tabular}
\end{table}
This procedure works if $\mu$ satisfies the condition 
$\Lambda_{QCD}\ll \mu \ll 1/a$ to avoid large higher-order perturbative corrections
and discretization errors. In the figure \ref{fig.1} we report the matrix
element of the renormalized
pseudoscalar operator in the chiral limit as a function of 
the renormalization scale for the
runs we have analysed. Data show that the discretization errors are within 
the statistical errors in the range $\mu a\simeq 1$ where the 
NP-renormalization is applied to renormalize the quark masses. 
All details of NP-renormalization will be given in a forthcoming paper 
\cite{finale}.
\vspace{-.3 cm}     
\section{RESULTS}
We have calculated the quark masses from the meson spectroscopy and from the
Ward Identity using different sets of quenched data with $\beta=6.0$, $6.2$ and
$6.4$ and using either the Wilson or the ``improved'' SW-Clover action. The
parameters used in each simulation are listed in Table 1 
and the main
results we have obtained are reported in Tables \ref{tab:qmas_spect} and
\ref{tab:qmas_wi}. The data at
$\beta=6.4$ have been used only for an exploratory study. The physical
volume and the time extension of the lattice are too small to be considered 
reliable.\\
The $\overline{MS}$ mass values in Table \ref{tab:qmas_wi} show 
that the non-perturbative approach allows a
determination of the quark masses with a smaller error than the standard
perturbative approach.  
In the $\beta$ range studied, there is {\it no}
statistical evidence for an "$a$" dependence of the quark masses for the Clover
action. For the Wilson action, a mild tendency in the quark masses to
decrease with increasing $\beta$ exists but an extrapolation to the continuum 
limit is not reliable. The $\overline{MS}$ mass values in 
Tables \ref{tab:qmas_spect} and \ref{tab:qmas_wi} show a very good
agreement between the quark masses extracted from the
spectroscopy and from the Ward Identity using the non perturbative
determinations of the renormalization constants, while the same comparison
using the perturbative values of $Z_A$ and $Z_P$ is much poorer 
\cite{finale,gupta}.  
We believe that the best estimates for the
strange and charm quark masses are obtained from the non-perturbative 
approach using Wilson and Clover data at $\beta=6.0$ and $\beta=6.2$.     
Our best
results are $m_s^{\overline{MS}}(2\; GeV)=(123\pm 4\pm 15)\; MeV$ and
$m_c^{\overline{MS}}(2\; GeV)=(1525\pm 40\pm 100)\; MeV$. The first error is
the statistical one and the second is the systematic error 
estimated extracting the quark masses from different mesons \cite{finale}.               
The values of quark masses we have
obtained are in good agreement with previous determinations
\cite{guido,mio1,vittorio,sesam}. 

\end{document}